\documentclass[aps,pre,twocolumn,groupedaddress]{revtex4}

\usepackage{subfigure}
\usepackage{graphicx, titlesec}
\usepackage{amssymb, amsmath}
\usepackage{natbib}
\usepackage{setspace}
\usepackage{acronym}

\begin{document}

\title{Sensitivity of Global Dynamics on the microscopic details of a network of dynamically coupled maps}

\author{Ad\`{e}le Peel}
\author{Henrik Jeldtoft Jensen}
\email[Author to whom correspondence should be addressed:\\]
{h.jensen@imperial.ac.uk} 
\homepage{http://www.ma.imperial.ac.uk/~hjjens/}
\affiliation{Department of Mathematics, Imperial College London, South Kensington Campus, SW7 2AZ, London, UK}

\date{30 April 2007}


\begin{abstract}
Here we analyze the behavior of dynamically coupled maps, based on
those introduced in a series of papers by Ito \& Kaneko (\emph{Phys.
Rev. Lett.}, 88, 2002 \& \emph{Phys. Rev. E}, 67, 2003). We show how
the microscopic coupling mechanism changes the behavior of the
system both by affecting the stability of fixed points and through a
more subtle effect in the crossover behavior between different
regions of the parameter space. This makes it necessary to choose
very carefully the exact manner in which one couples maps if they
are to be used as a general model of composite systems.
\end{abstract}

\maketitle

\section{\label{Introduction}Introduction}

It is generally accepted that the specific details of microscopic
couplings are irrelevant for the collective behavior of globally
coupled maps (GCMs) (see \cite{KanekosBook} and references therein).
There are a multitude of arbitrarily different ways in which maps
may be coupled together, with none being clearly and generically
better. We show here that the exact mechanism of coupling can lead
to distinct behaviors of the system. Thus making it necessary to
choose very carefully the exact manner in which one couples maps if
they are to be used as a general model for composite systems.

Synchronization in systems of coupled maps has been widely observed
\cite{Kaneko1990, Popovych_et_al_PhysRevE2001,
Popovych_et_al_IntJBifChaos2000, Cencini_Torcini2005, ItoKaneko2002,
ItoKaneko2003}. They are therefore of immense interest as models for
understanding synchronization phenomena as observed in the
real-world such as fireflies \cite{SyncInFireflyRef}, heart
pacemaker cells \cite{SyncInHeartRef} etc. However, many of the
models to date have their connections prescribed a priori and as
such they are fixed in time. This is in contrast to real-world
systems where both the units and the connections are dynamic
elements. It is therefore desirable to extend these coupled systems
so as to allow the interactions to evolve along with the node
dynamics.

Models of interacting iterative maps fall into two general classes:
The first consists of those systems where the maps are coupled prior
to the nonlinear transformation, as in

\begin{equation}
x_{n+1}^i = f\Big( (1- \epsilon)x_n^i + \epsilon \sum_j
h(x_n^j)\Big)
\end{equation}
\noindent which shall be referred to as \emph{internal coupling}
(see for example \cite{Popovych_et_al_IntJBifChaos2000,
ItoKaneko2002}).

Alternatively, the nonlinear transformation may be applied prior
to the coupling, as in

\begin{equation}\label{Equation: external coupling form}
x_{n+1}^i = (1 - \epsilon )f(x_n^i) + \epsilon \sum_j h(x_n^j)
\end{equation}

\noindent which shall be referred to as \emph{external coupling}
(see for example \cite{Popovych_et_al_PhysRevE2001, Kaneko1990,
AtmanaspacherFilkandScheingraber}).

In the case when $h(x)=f(x)$ for external coupling and $h(x)=x$ for
internal coupling, the systems show equivalent dynamics. The
behavior of most interest is that for sufficiently high coupling
strength, the node-states synchronize; often providing an example of
the surprising phenomenon that is synchronized chaos. We show here
that it is possible to switch between this synchronized chaotic
behavior to a synchronized stationary state depending on the way the
coupling is implemented, namely the functional form of $h(x)$.


\section{\label{section: results}Model Details}

The models studied here are based on those introduced by Ito \&
Kaneko in \cite{ItoKaneko2002, ItoKaneko2003}. When external
coupling is employed, the system is described by
\begin{equation}\label{Equation: externally coupled Ito Kaneko model}
x_{n+1}^i = (1-c)f(x_n^i) + c \sum_{j=1}^N w_n^{ij}h(x_n^j)
\end{equation}
\begin{equation}\label{Equation: link evolution}
w_{n+1}^{ij} = \frac{[1+\delta \cdot
g(x_n^i,x_n^j)]w_n^{ij}}{\sum_{j=1}^N[1+\delta \cdot
g(x_n^i,x_n^j)]w_n^{ij}}
\end{equation}
\begin{equation}\label{Equation: hebbian g}
g(x_n^i, x_n^j) = 1- 2|x_n^i - x_n^j|
\end{equation}

The functional form of $g$ was chosen so as to employ Hebbian
dynamics \cite{Hebb_Minireview}, although the results shown here are
qualitatively unchanged if anti-Hebbian dynamics are employed by
using $g(x_n^i, x_n^j)=|x_n^i-x_n^j|$ instead of Eq.
\eqref{Equation: hebbian g} (compare Figures \ref{figure:
bifurcation diag ext h(x)=f(x) Hebbian} and \ref{figure: bifurcation
diag ext h(x)=f(x) Anti-hebbian}). $\delta$ is a constant that
governs the plasticity of the network and is set to $0.1$ throughout
the present work.

The equivalent model with internal coupling is defined by the same
equations, only with Eq. \eqref{Equation: externally coupled Ito
Kaneko model} replaced by
\begin{equation}\label{Equation: internally coupled Ito Kaneko model}
x_{n+1}^i = f\Big[ (1-c)x_n^i + c \sum_{j=1}^N w_n^{ij}x_n^j \Big]
\end{equation}

As with many GCM studies, we use the logistic map, $f(x) = ax(1-x)$
as the underlying nonlinear dynamics. This is one of the simplest
functions that can display such distinct behavior, from
stationary-state, through periodic to chaotic. There has also been
recent evidence to suggest that functions with the form of the
logistic map may indeed be of direct relevance in neuroscience
\cite{Kuhn_Aertsen_Rotter2004}. We will consider two different forms
of $h(x)$,
\begin{equation}\label{h(x)=x}
    h(x) = x
\end{equation}
\begin{equation}\label{h(x)=f(x)}
    h(x) = f(x) = ax(1-x)
\end{equation}

\noindent When we use $h(x)=f(x)$ in the externally coupled system,
the model is similar to many \emph{externally coupled} systems in
the literature. Yet, this corresponds to interactions between the
units occurring instantaneously which is obviously not possible in
real systems. We therefore introduce a degree of causality to the
model by simply using $h(x)=x$ instead.

Each system has two basic parameters governing the dynamics: the
nonlinear parameter $a$ and the coupling strength $c$. This $a$-$c$
parameter space is known to have a series of regions of
qualitatively different behavior (see \cite{ItoKaneko2002,
ItoKaneko2003} for a detailed discussion of these). These regions
are present in both the internally and externally coupled systems
studied, although the boundaries occur at different actual values
and with different characteristics as will be shown in Section
\ref{Subsection: Phase Boundaries}.

\begin{figure}[!htbp]
\graphicspath{{External_WithFeedback/}} \vspace{5cm} \centering
\includegraphics[width=8.5cm]{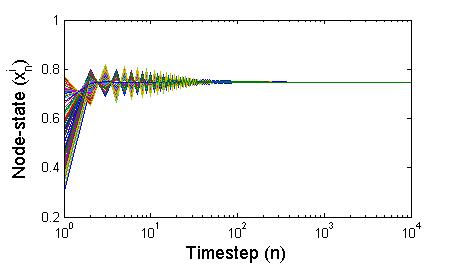}
\caption{Evolution of the node-states ($x^i$) for the external
coupling, $h(x)=x$ and $a=3.97$, $c=0.5$, $N=100$. The
time-evolution of each node-state, $x_n^i$, is represented by a
line on the graph. Thus giving $N$ lines on each graph. However,
when the nodes are synchronized these all lie on top of one
another so only one line may be seen. The nodes clearly
synchronize after $\thicksim 1000$ timesteps.}
     \label{figure: new model c=0.5 synchronous}
\end{figure}

\begin{figure}[!htbp]
\graphicspath{{External_WithoutDelays/}}\vspace{4cm}
\centering//\hspace{.5cm} 
\centering
\includegraphics[width=8.5cm]{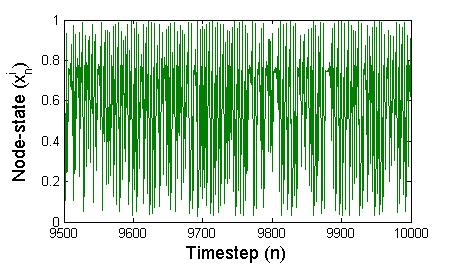}
\caption{Same as Fig. \ref{figure: new model c=0.5 synchronous}, but
for the model with internal coupling and $h(x)=f(x)$ for $a=3.97$,
$c=0.6$, $N=100$.}\label{figure: old model c=0.6 synchronous}
\end{figure}
\section{\label{Section: Results}Results}

As was found for systems with fixed couplings
\cite{AtmanaspacherFilkandScheingraber, Atay2006}, we find distinct
differences in the dynamics of the three systems: The system with
external coupling and $h(x) = x$ is found to enter a
stationary-state after an initial transient time, see Fig.
\ref{figure: new model c=0.5 synchronous}. This model could be
thought of as representing systems where interactions occur at
finite speed and there is therefore a time lag involved in the
updating. The value of the node-states in the stationary state
corresponds to the fixed point of the logistic map $\bar{x} =
\frac{a-1}{a}$.

In contrast, both the system with internal coupling and that with
external coupling but $h(x) = f(x)$ are never found to evolve into
this state. The behavior in the coherent region appears to vary at
random, see Fig. \ref{figure: old model c=0.6 synchronous}.

This distinction is not immediately obvious since the fixed point
of the logistic map, $\bar{x} = \frac{a-1}{a}$ is a fixed point of
both the externally and internally coupled map systems. It must
therefore be the stability of this fixed point that is crucially
dependant on the coupling mechanism employed.

In addition, the boundaries between the different phases of
parameter space occur at slightly different values of the parameters
and with different characteristics. This can be seen by comparing
Figs. \ref{Figure: sync transition internal}, \ref{Figure:sync
transition external no delay} and \ref{Figure:sync transition
external feedback}. Once again there is a stark difference between
the behavior of the externally coupled system with $h(x)=x$ and the
others. The trend with increasing system size for the external
system and $h(x)=x$ is for the boundary between ordered and coherent
behavior to occur at \emph{lower} coupling strength for larger
systems. This is the opposite of what happens with increasing system
size for both the internally coupled system and the externally
coupled system when $h(x)=f(x)$ where the boundary between ordered
and coherent behavior shifts to higher coupling strength for larger
systems.

\subsection{\label{Subsection:results:fixed point stability analytical}Fixed Point Stability: Analytical Results}

The stability of a fixed point of a dynamical system can be analyzed
through a simple perturbation analysis. Even to first order we can
see where the distinction between the externally coupled systems
with $h(x) = x$ and $h(x) = f(x)$ comes from. For a fixed point to
be stable, any perturbation, $\delta_n^i = (x_n^i - \bar{x})$ must
decay with time.

For the internally coupled system with $h(x)=x$, to first order we
have
\begin{equation}\label{Equation: internal, linear perturbation}
\delta_{n+1}^i\Big|_{h(x)=x} \simeq (2-a)(1-c)\delta_n^i +
(2-a)c\Big(\sum_j w_n^{ij}\delta_n^j\Big)
\end{equation}

\noindent For the externally coupled system with $h(x)=f(x)$, to
first order we have
\begin{equation}\label{Equation: external h(x)=f(x), linear perturbation}
\delta_{n+1}^i\Big|_{h(x)=f(x)} \simeq (2-a)(1-c)\delta_n^i +
(2-a)c\Big(\sum_j w_n^{ij}\delta_n^j\Big)
\end{equation}
\noindent For the externally coupled system with $h(x)=x$, to
first order we have
\begin{equation}\label{Equation: external h(x)=x, linear perturbation}
\delta_{n+1}^i\Big|_{h(x)=x} \simeq (2-a)(1-c)\delta_n^i +
c\Big(\sum_j w_n^{ij}\delta_n^j\Big)
\end{equation}

\noindent Consider Eq. \eqref{Equation: external h(x)=f(x), linear
perturbation}: The pre-factor $(2-a)$ to the second term plays a
vital role in the region we are interested in, $a>2$. Here,
$(2-a)<0$ so the first and second terms will be of the same sign and
therefore add constructively causing $\delta_n^i$ to oscillate about
the fixed point. Namely, that if $\delta_n^i<0$ then
$\delta_{n+1}^i>0$. Whereas if $\delta_n^i>0$ then
$\delta_{n+1}^i<0$.

In contrast, for the system with $h(x)=x$, Eq. \eqref{Equation:
external h(x)=x, linear perturbation}, there is no negative
pre-factor to the second term. Thus, the two terms will be of
opposite sign and so add de-constructively. This cancelation of one
another is what leads to the reduction in $\delta_n^i$ and therefore
the stability of the fixed point.

We have analyzed these terms through numerical simulations and found
the results (not shown here) to confirm this conjecture of the terms
canceling one another for the external system with $h(x)=x$. For
this system, the two terms were found to be of comparable amplitude
and opposite sign and thus the fixed point of the system is stable
to perturbations.

\subsection{\label{Subsection:results:fixed point stability numerical}Fixed Point Stability: Numerical Results}

\begin{figure}[!htbp]
\graphicspath{{External_WithFeedback/}} \vspace{4cm} \centering
\includegraphics[width=8cm]{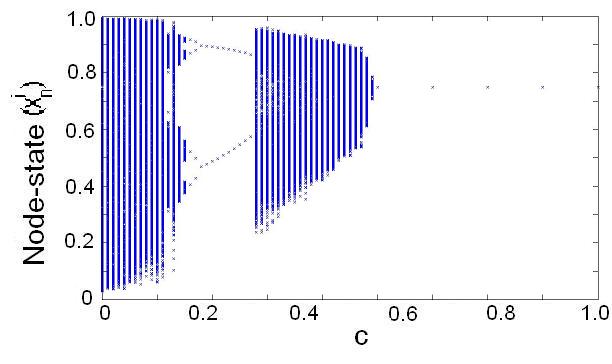} \caption{A bifurcation
diagram for the system with external coupling and $h(x)$$=$$x$. The
plot shows the behavior of all nodes for 100 time steps at various
$c$-values for fixed $a=3.97$. For each time step, each $x_n^i$ is
plotted. For higher coupling values when the system is synchronized,
these will all coincide and there will therefore appear to be less
points plotted.}\label{figure: bifurcation diag ext h(x)=x}
\end{figure}

\begin{figure}[!htbp]
\graphicspath{{External_WithoutDelays/}}\vspace{4cm}
\centering
\centering
\includegraphics[width=8cm]{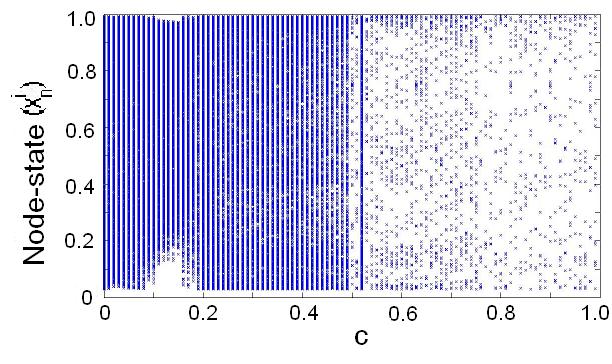}
\caption{A bifurcation diagram for the system with external coupling
and $h(x)$$=$$f(x)$. The plot shows the behavior of all nodes for
100 time steps at various $c$-values for fixed $a=3.97$. The points
become markedly less dense after $c$ $\thicksim$ $0.5$ because the
system is synchronized in this region. Thus, each time-step results
in only one point on the graph as opposed to 100 (one for each node)
as is the case when the system is unsynchronized.}\label{figure:
bifurcation diag ext h(x)=f(x) Hebbian} 
\end{figure}

\begin{figure}[!htbp]
\graphicspath{{Internal_WithoutTimeDelays/}} \vspace{4cm} \centering
\centering
\includegraphics[width=8cm]{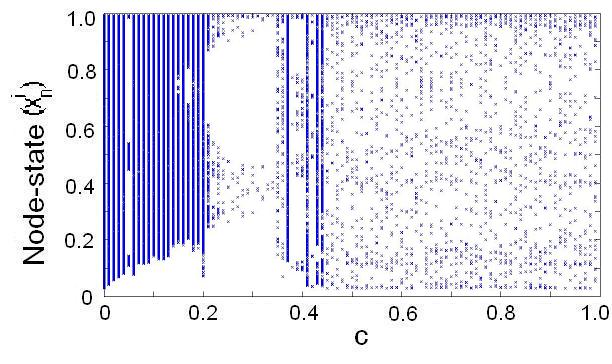} \caption{A
bifurcation diagram for the system with internal coupling. The plot
shows the behavior of all nodes for 100 time steps at various
$c$-values for fixed $a=3.97$.}\label{figure: bifurcation diag int
without delay}
\end{figure}

\begin{figure}[!htbp]
\vspace{4cm} \centering
\graphicspath{{External_WithoutDelays/Anti-Hebbian/}}
\centering
\includegraphics[width=8cm]{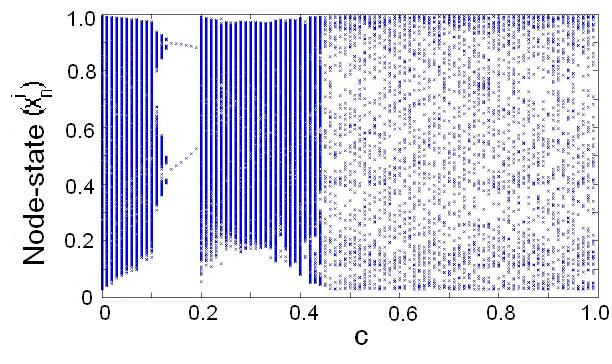}
\caption{The same as Figure \ref{figure: bifurcation diag ext
h(x)=f(x) Hebbian} but employing Anti-Hebbian dynamics in the
evolution of the connection strengths. The plot shows the behavior
of all nodes for 70 time steps.}\label{figure: bifurcation diag ext
h(x)=f(x) Anti-hebbian}
\end{figure}

Through numerical simulations of the three systems we have been able
to characterize the behavior displayed by each system. Through
plotting bifurcation diagrams, we can see how this behavior is
dependant upon the coupling strength parameter, $c$.

Fig. \ref{figure: bifurcation diag ext h(x)=x} shows us that the
externally coupled system with $h(x)=x$ is in fact stable on the
fixed point for a range of coupling strengths ($c$-values). This is
in contrast to the other systems whose bifurcation diagrams show
that the systems do not evolve onto a fixed point at any coupling
strength, see Figs. \ref{figure: bifurcation diag ext h(x)=f(x)
Hebbian} and \ref{figure: bifurcation diag int without delay}.

When the system is reduced to the standard GCM with all-to-all
couplings that are constant in time, these distinguishing behaviors
of the systems are effectively unchanged. Namely, that the
externally coupled system with $h(x)=x$ has a stable fixed point
whereas the other systems do not. This result should be expected
since the stability as explained through linear perturbation
analysis is \emph{not a result of the network topology}.
\subsection{\label{Subsection: Phase Boundaries}Phase Boundaries}

As was shown by Ito \& Kaneko in \cite{ItoKaneko2003} the $a$-$c$
parameter space consists of distinct regions which are qualitatively
different. The regions are essentially those first described in
\cite{Kaneko1990}; namely that for sufficiently high coupling
strength there is a region of $a$-$c$ parameter space where all
nodes synchronize, displaying \emph{coherent} behavior. For lower
coupling strength, this global synchronization is lost and the
system splits into distinct clusters, this region of $a$-$c$
parameter space is known as the \emph{ordered} region. For yet lower
values of the coupling strength there is a \emph{disordered} region,
where there is no synchronous behavior between any pair of nodes.

We have found that as one crosses the boundary from ordered to
coherent regions, the different systems show qualitatively different
transitions. This can be seen by looking at the behavior of
$\sigma^2$ as we cross the boundary by changing $c$ at fixed
$a$-value. Where, $\sigma^2$ is as defined in
\cite{MartiMasoller05}. Namely,
\begin{equation}
\sigma^2 = \frac{1}{N}\big<\sum_i[x_n^i-<x_n>]^2\big>_t
\end{equation}

\noindent where $<x_n>$ denotes the average over all node-states
$x^i$ at timestep $n$ and $\big<\dotsi\big>_t$ denotes the average
over all timesteps.

\begin{figure}[!htbp]
\vspace{4cm} \centering
\graphicspath{{Internal_WithoutTimeDelays/}} \centering
\includegraphics[width=8.5cm]{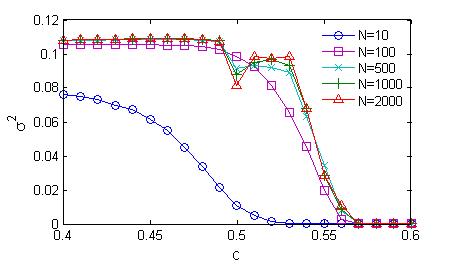}
\caption{Here we see the transition to synchronization and how it
changes with the system size for the internally coupled system. For
ease of comparison, we re-scale the data and plot $\sigma^2$ versus
the coupling strength $c$. This data is the average over 1000 random
initial conditions. Here we show the transition for different system
sizes, $N=10$ ($\square$), $N=100$ ($\bigcirc$), $N=500$ (x),
$N=1000$ (+) and $N=2000$ ($\bigtriangleup$).}
     \label{Figure: sync transition internal}
\end{figure}

\begin{figure}[!htbp]
\graphicspath{{External_WithoutDelays/}} \centering \vspace{4.5cm}
\includegraphics[width=8.5cm]{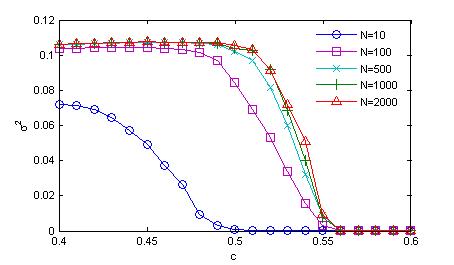}
\caption{Same as Figure \ref{Figure: sync transition internal} but
for the externally coupled system with $h(x)=f(x)$.}
\label{Figure:sync transition external no delay}
\end{figure}

\begin{figure}[!htbp]
\vspace{4.5cm} \centering \graphicspath{{External_WithFeedback/}}
\includegraphics[width=8.5cm]{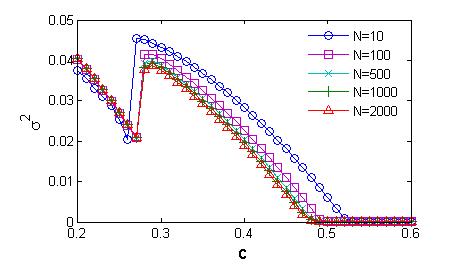}
\caption{Same as Figure \ref{Figure: sync transition internal} but
for the externally coupled system with
$h(x)=x$.}\label{Figure:sync transition external feedback}
\end{figure}

Figs. \ref{Figure: sync transition internal}, \ref{Figure:sync
transition external no delay} \& \ref{Figure:sync transition
external feedback} show that there is a fundamental difference in
the crossover from ordered (distinct synchronized clusters) to
coherent (one giant synchronized cluster) regions between the three
systems. The externally coupled system with $h(x)=f(x)$ and the
internally coupled system have qualitatively similar behavior in
$\sigma^2$; with increasing system size the transition moves to
higher $c$-values and the sharpness increases. There is however one
local defining feature for the internally coupled system, at
$c=0.5$: there is a localized decrease in $\sigma^2$. In contrast to
either of these two systems, for the externally coupled system with
$h(x)=x$, the transition moves to \emph{lower} $c$-values for larger
system sizes. If these results are extrapolated to the limit of
infinite system size, this suggests that only the externally coupled
system with $h(x)=x$ would be able to synchronize globally.

One feature that is common to all three systems however is the fact
that the nature of the transition from one phase to another (ordered
to coherent) is fundamentally different to other types of
\emph{phase transition}. A hallmark of phase transitions in
equilibrium statistical mechanics is a discontinuity in the order
parameter, in the limit of an infinite system size. Numerically,
this can be seen by a sharpening of the transition with increasing
system size. However, here the sharpening as seen in Figs.
\ref{Figure: sync transition internal} \& \ref{Figure:sync
transition external no delay} possesses a subtle distinction since
it is a probabilistic measure of the size of the basin of attraction
for the coherent state.

For a given point in the $a$-$c$ parameter space and set of initial
conditions, the system will evolve into the coherent state say. For
an arbitrarily small change in the initial conditions we see that
the system is attracted to the ordered state of several clusters.
This apparent \emph{riddling} of the basin of attraction of the
coherent region is ordinarily not observed in standard phase
transitions and gives rise to a broad and structured boundary
between the coherent and ordered phases in the $a$-$c$ parameter
space. Indeed, such a riddling has been found in other coupled map
systems \cite{MaistrenkoMaistrenkoPopovychMosekilde98}. It has also
been long since conjectured that riddled basins of attraction could
be extremely common in natural systems \cite{SommererOtt93}. In such
a scenario, the parameter space phase diagrams can be misleading
since they give no indication of this broad parameter region that is
neither wholly in one phase nor the other.
\section{\label{Conclusions}Conclusions}

We have shown that the microscopic details of dynamic couplings have
a dramatic effect on the macroscopic dynamics displayed by systems
of globally coupled maps. Systems with asynchronous coupling are, in
some cases, able to stabilize a fixed point of the underlying map.
This stabilizing effect of coupling shown here is the same as was
previously shown for coupled map lattices with fixed un-weighted
coupling \cite{AtmanaspacherFilkandScheingraber,Atay2006}. Since the
stabilizing effect is found when there is a finite speed at which
the interaction between nodes is transmitted, this does call into
question the applicability of other coupled systems as prototype
models for synchronized chaos of real world systems.

We must also question what, if any, significance may be attached to
the form of coupling that must be employed in order to gain
synchronized chaos, namely the internal coupling or synchronously
updated external coupling. If these models are to be viewed as
prototype models of real world synchronization phenomena, we may use
this result to gain insight into the mechanisms governing real
systems.

Further, we have shown that this stabilizing of fixed points is not
the only macroscopic difference between the systems. The crossover
from one region of parameter space to another has been shown to be
highly variable between the different systems studied. Although all
systems show non-trivial mechanisms as one crosses from one phase to
another. The boundary maintains a finite width with a great deal of
structure to it in all three systems studied.
\section{\label{Acknowledgements}Acknowledgements}
The authors would like to thank Gil Benkoe for useful discussions
and Kunihiko Kaneko for helpful email correspondence. Adele Peel
gratefully acknowledges the Engineering and Physical Sciences
Research Council (EPSRC) for her Ph.D. studentship.
\bibliographystyle{unsrt}
\bibliography{bibliography2}

\begin{thebibliography}{10}

\bibitem{ItoKaneko2002}
J.~Ito and K.~Kaneko.
\newblock Spontaneous structure formation in a network of chaotic units with
  variable connection strengths.
\newblock {\em Phys. Rev. Lett.}, 88:028701, 2002.

\bibitem{ItoKaneko2003}
J.~Ito and K.~Kaneko.
\newblock Spontaneous structure formation in a network of dynamic elements.
\newblock {\em Phys. Rev. E}, 67:046226 2003.

\bibitem{KanekosBook}
K.~Kaneko and I.~Tsuda.
\newblock {\em Complex Systems: Chaos and Beyond}.
\newblock Springer-Verlag, 2000.

\bibitem{Kaneko1990}
K.~Kaneko.
\newblock Clustering, coding, switching, hierarchical ordering and control in a
  network of chaotic elements.
\newblock {\em Physica D}, 41:137--172, 1990.

\bibitem{Popovych_et_al_PhysRevE2001}
O.~Popovych, Yu. Maistrenko, and E.~Mosekilde.
\newblock Loss of coherence in a system of globally coupled maps.
\newblock {\em Phys. Rev. E}, 64:026205, 2001.

\bibitem{Popovych_et_al_IntJBifChaos2000}
Yu. Maistrenko, O.~Popovych, and M.~Hasler.
\newblock On strong and weak chaotic partial synhronization.
\newblock {\em Int. Jour. Bif. and Chaos}, 10(1):179--203, 2000.

\bibitem{Cencini_Torcini2005}
M.~Cencini and A.~Torcini.
\newblock Nonlinearly driven transverse synchronization in coupled chaotic
  systems.
\newblock {\em Physica D}, 208:191--208, 2005.

\bibitem{SyncInFireflyRef}
J.~M. Buck.
\newblock Synchronous rythmic flashing of fireflies.
\newblock {\em Quarterly Review of Biology}, 13:301--314, 1938.

\bibitem{SyncInHeartRef}
D.~C. Michaels, E.~P. Matyas, and J.~Jalife.
\newblock Mechansims of sinoatrial pacemaker synchronization: A new hypothesis.
\newblock {\em Circulation Research}, 61:704--714, 1987.

\bibitem{AtmanaspacherFilkandScheingraber}
H.~Atmanaspacher, T.~Filk, and H.~Scheingraber.
\newblock Stability analysis of coupled map lattices at locally unstable fixed
  points.
\newblock {\em Eur. Phys. J. B}, 44:229 -- 239, 2005.

\bibitem{Hebb_Minireview}
T.~J. Sejnowski.
\newblock The book of hebb.
\newblock {\em Neuron}, 24:773--776, 1999.

\bibitem{Kuhn_Aertsen_Rotter2004}
A.~Kuhn, A.~Aertsen, and S.~Rotter.
\newblock Neuronal integration of synaptic input in the fluctuation-driven
  regime.
\newblock {\em J. Neuroscience}, 24:2345, 2004.

\bibitem{Atay2006}
F.~M. Atay and O.~Karabacak.
\newblock Stability of coupled map networks with delays.
\newblock {\em SIAM Journal of Applied Dynamical Systems}, 5:508--527, 2006.

\bibitem{MartiMasoller05}
A.C. Marti and C.Masoller.
\newblock Random delays and the synchronization of chaotic maps.
\newblock {\em Phys. Rev. Lett.}, 94:134102, 2005.

\bibitem{MaistrenkoMaistrenkoPopovychMosekilde98}
Yu.L. Maistrenko, V.L. Maistrenko, A.~Popovych, and E.~Mosekilde.
\newblock Role of the absorbing area in chaotic synchronization.
\newblock {\em Phys. Rev. Lett.}, 80:1638 -- 1641, 1998.

\bibitem{SommererOtt93}
J.C. Sommerer and E.~Ott.
\newblock A physical system with qualitatively uncertain dynamics.
\newblock {\em Nature}, 365:138 -- 140, 1993.

\end{thebibliography}
\end{document}